\RequirePackage{lineno}
\documentclass[aps,prl,twocolumn,numerical,superscriptaddress,nofootinbib,showpacs,showkeys,longbibliography]{revtex4-1}

\usepackage{lineno}
\usepackage{graphicx}
\makeatletter
\renewcommand{\p@subsection}{}
\renewcommand{\p@subsubsection}{}
\makeatother

\begin{document}

\newcommand{\Nch}{N_{\rm ch}}
\newcommand{\dg}{\Delta\gamma}
\newcommand{\R}{R_{\Psi_{2}}}
\newcommand{\dS}{\Delta S}
\newcommand{\sigR}{\sigma_{\R}}
\newcommand{\mean}[1]{\langle#1\rangle}

%\linespread{1.6}
\title{Rebuttal to: ``Comment on Scaling properties of background- and chiral-magnetically-driven charge separation in heavy ion collisions at $\sqrt{s}_{\rm NN}=200$~GeV''}

\author{Roy~A.~Lacey} 
\email{Roy.Lacey@stonybrook.edu}
\affiliation{Depts. of Chemistry \& Physics, Stony Brook University, Stony Brook, New York 11794, USA}

\author{Niseem~Magdy} 
%\email{niseemm@gmail.com}
\affiliation{Department of Chemistry, Stony Brook University, Stony Brook, New York 11794, USA}
%--------------------------------------------------------------------------------------------------------------------------------------------
\date{\today} %this is useful in drafting stage
%\draftversion{12}

%------------------------------------------------------------------------------------------------%

\begin{abstract}
Recently, F. Wang commented \cite{Wang:2022fth} on our work -- ``Scaling properties of background- and chiral-magnetically-driven charge separation in heavy ion collisions at $\sqrt{s}_{\rm NN}~=~200$ GeV'' \cite{Lacey:2022baf}  – and made several claims to support his conclusion that the results in Ref. \cite{Lacey:2022baf} are fallacious. His conclusion and claims are not only incorrect; they show a fundamental disconnect with the rudiments of the $R_{\Psi_2}(\Delta S)$ correlator. This rebuttal addresses the root misconception responsible for Wang's false claims.
\end{abstract}
\maketitle

%------------------------------------------------------------------------------------------------%

In the recent publication of the STAR Collaboration's isobar data~\cite{STAR:2021mii}, the ratios between Ru+Ru and Zr+Zr collisions of the observables $\dg/v_2$ (the azimuthal correlator $\dg$~\cite{Voloshin:2004vk} divided by the elliptic flow coefficient $v_2$) and $1/\sigR$ (the inverse width of the $\R(\dS)$ distribution~\cite{Magdy:2017yje}) were reported. Both correlators indicated ratios ($R_{\rm Ru/Zr}$) less than unity but more significant deviations from unity for $\dg/v_2$. Because these ratios are different from the predefined  values $R_{\rm Ru/Zr} > 1$ expected for the chiral magnetic effect (CME)~\cite{STAR:2021mii}, they precipitated the conclusion that the signal difference (between the isobars) obtained in the  STAR blind analysis is incompatible with the presence of a CME signal. This conclusion is, of course, predicated on the notion that the background difference between Ru+Ru and Zr+Zr is negligible and the correlators are sensitive to the small-signal difference expected~\cite{Magdy:2020wiu}. A non-negligible background difference influencing $R_{\rm Ru/Zr}$ was reported in Ref.~\cite{STAR:2021mii} and several post-blind analyzes' have attempted to evaluate its consequence on the ratios for $\sigR^{-2}$~\cite{Lacey:2022baf} and $\dg/v_2$ \cite{STAR:2021mii,Tribedyqm,Kharzeev:2022hqz,Lacey:2022plw}. This rebuttal addresses the essential questions raised by F. Wang in his comment \cite{Wang:2022fth} to the post-blind results we reported in Ref.~\cite{Lacey:2022baf}.

A central underpinning to Wang's critique~\cite{Wang:2022fth} of our work in Ref.~\cite{Lacey:2022baf} is the misconception that $\sigR^{-2}~\approx~\Nch\dg~\propto~v_2$  \cite{STAR:2021mii,Wang:2022fth}, where $\Nch$ is the charged particle multiplicity. In the following, we debunk this misconception by showing that the $R_{\Psi_2}(\Delta S)$ correlator for background-driven charge separation is  only sensitive to the charge-dependent non-flow background~\cite{Magdy:2017yje,Magdy:2018lwk,Sun:2018idn}, and $\sigR^{-2}~\propto ~1/\Nch$.

The $R_{\Psi_2}(\Delta S)$ correlator measures charge separation relative to the $\Psi_2$ plane via the ratios:
\begin{eqnarray}
\label{eq1}
R_{\Psi_2}(\Delta S) = C_{\Psi_2}(\Delta S)/C_{\Psi_2}^{\perp}(\Delta S), \\ 
C_{\Psi_2}(\Delta S)=\frac{N(\Delta S_{\rm Real})}{N(\Delta S_{\rm Shuffled})}, C^{\perp}_{\Psi_2}(\Delta S)=\frac{N(\Delta S^{\perp}_{\rm Real})}{N(\Delta S^{\perp}_{\rm Shuffled})},
\label{eq2}
\end{eqnarray}
where $C_{\Psi_2}(\Delta S)$ and $C_{\Psi_2}^{\perp}(\Delta S)$ are correlation functions that quantify charge separation $\Delta S$, approximately parallel and perpendicular (respectively) to the $\vec{B}$-field. The charge shuffling procedure employed in constructing these correlation functions ensures identical properties for their numerator and denominator, except for the charge-dependent correlations, which are of interest~\cite{Magdy:2017yje,Magdy:2018lwk}. 

The inverse variance $\sigma^{-2}_{R_{\Psi_{2}}}$ of the $R_{\Psi_2}(\Delta S)$ distributions quantify the charge separation~\cite{Magdy:2017yje,Magdy:2018lwk,Shi:2019wzi} as:
\begin{eqnarray}
\label{eq3}
\sigma^{-2}_{R_{\Psi_{2}}} = \sigma_{\Delta S_{\rm Real}}^{-2} - \sigma_{\Delta S_{\rm  Shuffled}}^{-2} - \sigma_{\Delta S^{\perp}_{\rm Real}}^{-2} + \sigma_{\Delta S^{\perp}_{\rm Shuffled}}^{-2}  \\
 = \left[ \sigma_{\Delta S}^{-2}  - \sigma_{\Delta S^{\perp}}^{-2} \right]_{\rm Real} - \left[ \sigma_{\Delta S}^{-2}  - \sigma_{\Delta S^{\perp}}^{-2}\right]_{\rm Shuffled}, 
\label{eq4}
\end{eqnarray}
indicating that $\sigma^{-2}_{R_{\Psi_{2}}}$  is the difference between the inverse variances for the distributions of Real and Shuffled events.  
\begin{figure}[t]
  \includegraphics[width=1.0\linewidth]{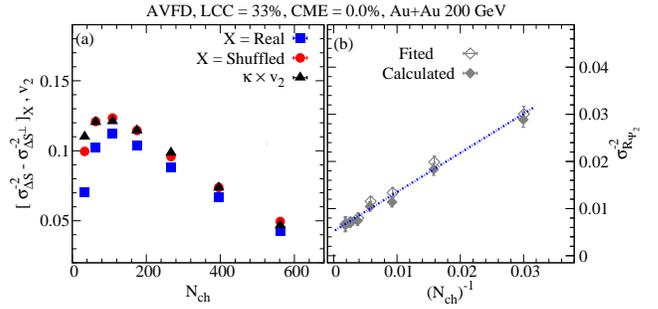}
  \caption{Comparison of the $\Nch$-dependence of $v_2$ and the inverse variances for Real and Shuffled events (cf. Eq.~\ref{eq4}) obtained for background-driven chrage separation simulated with the AVFD model (a). The open and closed symbols in panel (b) compare $\sigma^{-2}_{R_{\Psi_{2}}}$ vs. $1/\Nch$ for fits to the $R_{\Psi_2}(\Delta S)$ distributions (Eq.~\ref{eq1})  and the inverse variances extracted via Eq.~\ref{eq4}.  
	}
  \label{fig1}
\end{figure}
This is illustrated in Fig.~\ref{fig1} for events simulated with the anomalous viscous fluid dynamics (AVFD) model~\cite{Shi:2017cpu} for background-driven charge separation; these results incorporate the requisite corrections for number fluctuations and event plane resolution~\cite{Magdy:2017yje,Magdy:2018lwk}. Fig.~\ref{fig1}(a) shows that  $v_2$ is proportional to the shuffled term in Eq.~\ref{eq4}, indicating that the difference between the Real and Shuffled terms [Eq.~\ref{eq4}] reflects only the influence of the charge-dependent non-flow correlations. The well-known $1/\Nch$ dependence of such correlations is made more transparent in Fig.~\ref{fig1}(b), which shows that $\sigma^{-2}_{R_{\Psi_{2}}} \propto 1/\Nch$; this dependence is similar for both isobars but with different  $\Nch$ values. Wang et al.~\cite{STAR:2021mii,Wang:2022fth,Feng:2020cgf,Choudhury:2021jwd} have been repeatedly made aware of these facts to no avail.

In Ref.~\cite{Lacey:2022baf}, the $\sigma^{-2}_{R_{\Psi_{2}}}$ values extracted for background and signal + background at a given centrality, were checked to establish their sensitivity to variations in the magnitude of the anisotropic flow coefficient $v_2$, using event-shape selection via fractional cuts on the distribution of the magnitude of the $q_2$ flow vector \cite{Schukraft:2012ah}. The checks indicated that, while $v_2$ shows a sizable increase with $q_2$, the corresponding $\sigma^{-2}_{R_{\Psi_{2}}}$ values are insensitive to $q_2$ regardless of background or signal + background. Similar patterns were observed for the Isobar data reported in Ref.~\cite{STAR:2021mii}. 

In contrast to Wang's stated confusion about the $q_2$ dependence of $\sigma^{-2}_{R_{\Psi_{2}}}$~\cite{STAR:2021mii,Wang:2022fth}, it is straightforward to see that the observed insensitivity stems from a cancellation which results from the difference between the Real and Shuffled terms in Eq.~\ref{eq4} (cf. Fig.~\ref{fig1}). Note, however, that $q_2$ selection methods which result in a small $\Nch$  bias, especially for large $q_2$ \cite{STAR:2021mii}, could lead to small modifications to the insensitivity trend. It is straightforward to implement a methodological change that prevents a possible $\Nch$  bias. Wang et al.~\cite{STAR:2021mii,Wang:2022fth} has persistently ignored these facts.

In summary, we have shown that $\sigma^{-2}_{R_{\Psi_{2}}} \propto 1/\Nch$ which debunks the the root  claim by Wang et al. that $\sigR^{-2}~\approx~\Nch\dg~\propto~v_2$~\cite{STAR:2021mii,Wang:2022fth,Feng:2020cgf,Choudhury:2021jwd}. This falsification renders all direct and collateral inferences in Wang's comment~\cite{Wang:2022fth} false. 
%'s 
\vspace{0.3cm}

This research is supported by the US Department of Energy, Office of Science, Office of Nuclear Physics, 
under contracts DE-FG02-87ER40331.A008

\bibliography{lpvpub}

\end{document}